# Comoving to expansion Newtonian potential of galaxies and clusters instead of dark matter

Zahid Zakir[1]

**Abstract**

Stretching of the Newtonian potential (NP) at early epochs is investigated and it is shown that observed effects, usually ascribed to a dark matter, can by explained by such stretching only. Increasing by time radius of the gravitationally-bound region (GBR) and conservation of gravitational energy lead to a new scenario in which values of NP in expanding volume are maintained, while in physical volume are stretched. Really, the energy conservation in expanding volume requires for NP values to be comoving to the expanding shells. In addition, the radius of gravitationally-bound region increases by time due to decreasing of expansion velocity and different shells around galaxy cease expansion at different times. Thus, as far a shell placed from galaxy, as longer it was expanded and thickened, while potential difference on its boundaries remained unchanged. This shifts the values of NP around galaxy proportional to the distance r and, as the result, the gravitational acceleration, from NP's $1/r^2$ dependence, turned to $1/r$ dependence, as for centrifugal acceleration. This fact naturally explains the known empirical facts, such as flatness of rotation curves and velocity-mass relationships for galaxies and velocity dispersion in clusters.

Metric expansion of space means that the expanding volume element $V(t) \sim a^3(t)$, for example interior of a thin dust shell, contains more units of physical volume than the same volume at earlier moment $V(t_0)$. The increased volume contains the same galaxies, but now their gravitational fields extend to the volume larger to $\delta V = V(t) - V(t_0)$.

When galaxies would displaced in static space, the increased distance between neighbors would occupied by the potential of each galaxy existed before at the same physical distance from it. In expanding space the question is non-trivial and there are two possibilities for the values of potential – or they maintain at the same physical distances and the potential does not change, or they comove to receding sample particles and the potential becomes "stretched".

---
[1] *Centre for Theoretical Physics and Astrophysics, 11a, Sayram Str., 100170 Tashkent, Uzbekistan*,
zahidzakir@theor-phys.org

Here as a criterion of a choice appears the energy of the gravitational field which in general case should be conserved. Therefore, for application of gravity theory to intergalactic expanding space one must answer to question about what happens by the energy of gravitational field of galaxies in the expanding volume. The answer is reduced to choosing one of two possibilities, when energy of the field:

(*a*) conserves in the physical volume element, but then it grows (on the module) in the expanding volume element $V(t)$;

(*b*) conserves in the expanding volume element $V(t)$, but then it decreases (on the module) in the physical volume element.

Simplest treatment, accepted as a standard one, is based on the first version (*a*) by direct extrapolation of NP to galaxy scale distances [1]. In this case for a rotating galaxy of mass *M* the balance of centrifugal and gravitational forces $v_c^2 / r = GM(r)/r^2$ gives for the rotation velocity the distance dependence $v_c(r) \sim r^{-1/2}$. But, for spiral galaxies observations [1] show the flatness of rotation curves $v_c(r) \sim const$. This requires supposing of non-baryonic dark matter haloes distributed as $M(r) \sim r$. Thus, the standard treatment requires new hypotheses and, in addition, is unsatisfactory theoretically since the field energy grows by time in any expanding volume element, including the entire universe too.

In the present paper a new treatment based the second answer (*b*) is formulated. In it the conservation of field's energy in any expanding volume element leads to its conservation in entire universe too, and thus at least theoretically this approach is more preferable.

Other new feature of this treatment is appropriate accounting of the existence and evolution of radius $r_c$ of GBR of galaxy, at interior of which the field is static [2]. This radius corresponds to a distance where particle's gravitational potential energy in the galaxy's field compensates the kinetic energy corresponding to the receding velocity $v_H(r) = H r$, where $H(a) = \dot{a}/a$ - Hubble parametre, which gives:

$$H^2 r_c^2 / 2 = GM / r_c, \quad r_c = (2GM / H^2)^{1/3}. \tag{1}$$

The receding velocity outside $r_c$ restores not immediately, but increases as:

$$v_{H(s)}(r) = H_s r, \quad H_s(r) = H \cdot (1 - r^3 / r_c^3)^{1/2}. \tag{2}$$

Taking $H = h \times 100 km \times s^{-1} \times Mpc$, at the present value $h = h_0 = 0.7$ and masses $M = (10^{10}; 10^{11}; 10^{12}; 10^{13}) M_\odot$ one has $r_{c(0)} = (0.26; 0.56; 1.2; 2.6) Mpc$ accordingly, while at early epochs with $h = 10$ one has $r_{c(0)} = (0.044; 0.095; 0.20; 0.44) Mpc$. Thus, in the galaxies formation epoch $r_c$ had a value about a size of galaxies, and then had increased by time as $r_c \sim H^{-2/3}$ due to decreasing of $H$ by growing of the scale factor $a(t)$.

Let at early epochs a spatial shell, around a galaxy, of initial radius $r$ and thickness $\Delta r$ participates at the expansion. After some time $r_c$ exceeds shell's radius and it becomes static by reaching fixed radius $r \to r'$ and thickness $\Delta r \to \Delta r$. Such expanded and then successively stabilized shells becomes located as farther from their initial radii $r$, as more time each of them was expanded and thickened. In the first linear approximation the thickening is proportional to the distance:

$$r \to r', \quad \Delta r \to \Delta r = (r'/r_c) \cdot \Delta r'. \tag{3}$$

In such spatial layers around galaxy, which expanded and then layer by layer stabilized, the gravitational field occurred deformed due to conservation of field's energy in any layer.

Let's consider the gravitational potential in these shells. In each of them the field deforms differently - as long time a shell was expanded, as lower felled the field's energy density in this shell with respect to its initial value. The energy density $W(r) = -(\nabla \phi_g)^2 / 8\pi G$ of NP $\phi_g = -GM/r$ is $W(r) = -GM^2/8\pi r^4$ and the field's energy in spherical shell $r_2 - r_1$ is $U(r_2 - r_1) = [\phi_g(r_1) - \phi_g(r_2)]M/2$. At expansion of a homogeneous dust ball, the energy conservation of the field requires that the difference of the gravitational potential between shells does not change and the values of the potential *comove* to these shell. Thus, if the concept of a relativistic field required an introduction of the *retarded* potentials, the concept of such field in expanding space requires the introduction of *the comoving potentials* for the gravity theory.

In a shell having initial thickness $\Delta r$, which later expanded and became thicker as in (3), the field's energy conservation condition at such shifting and thickening takes the form:

$$\phi_g(r + \Delta r) - \phi_g(r) = \phi_g\left(r' + (r'/r_c) \cdot \Delta r'\right) - \phi_g(r'). \tag{4}$$

i.e. during expansion the potential difference on boundaries of the shell does not change. The conserving gravitational energy in the shell, thus, is given by:

$$\Delta U = -\frac{GM}{r^2}\Delta r = -\frac{GM}{r'^2} \cdot \frac{r'}{r_s}\Delta r' = -\frac{GM}{r' r_s}\Delta r', \tag{5}$$

from which we find the modified gravitational acceleration

$$w_{g(s)}(r') = -\frac{GM}{r' r_s}. \tag{6}$$

Thus, the Newtonian potential, stretched during expansion at early epochs, turns to the comoving potential $\phi_g(r) \to \phi_g(r')$ with corresponding modified gravitational acceleration $w_g(r) \to w_{g(s)}(r')$.

As the result, at such stretched potential the balance of centrifugal and gravitational forces leads to a constant rotation velocity of galaxy in the region $r > r_c$:

$$\frac{v_c^2}{r'} = \frac{GM}{r_c r'}, \quad v_c^2 = \frac{GM}{r_c} = const. \tag{7}$$

It explains the flatness of the rotation curves of galaxies [1] and allows to define $r_c$ from data for masses and rotation velocities $r_c = GM/v_c^2$. For $r$ observations show $r_c \ll r$, which is natural since an initial value of $r_c$ was slightly smaller than the present radius of the galaxy. Bounding of galaxies into clusters with switching off from the expansion flow, after which only clusters became mutually receding, occurred at an epoch when the value $r_c$ became more than a half of average distance between neighbor galaxies in clusters.

From Eq. (7) it follows a relationship between the rotation velocity and mass of galaxy - by inserting $r_c$ from (1) into (7) we obtain:

$$v_c^2 = (GMH_c)^{2/3}, \quad v_c^3 = (GH_c)M. \tag{8}$$

Here $H_c$ is the value of Hubble parametre at the epoch when the galaxy's field had formed, and $M$ is baryonic mass of the galaxy inside $r_c$.

It is known the empirical formula - the baryonic Tully-Fisher (BTF) relationship $M_b = \alpha v^\beta$, where $\alpha = const.$ and $\beta \simeq 3 \div 4$ at different estimations of mass. The observable baryonic mass of galaxy $M_b$, including mass outside $r_c$ too, exceeds $M$ and, consequently, obtained in (8) theoretical dependence $M \sim v^3$ with $\beta = 3$ really gives lower limit for $\beta$.

But, for rough estimation of the parameters at early epochs, we can consider also upper limit with $\beta = 4$, when Eq. (8) with the expression for mass $M = \alpha v^4$ gives $H_c = (G\alpha v)^{-1}$. The observing rotation velocity $v \sim 200 km \times s^{-1}$ and the empirical value $\alpha \simeq 50 M_\odot \times km^{-4} \times s^4$ then lead to the estimations $H_c \sim 20 km \times s^{-1} \times kpc^{-1}$ and $r_c \sim 5 \div 10 kpc$, which are satisfactory, since this distance coincides by starting point of flatness region in observing rotation curves. If we exclude from the relationships velocity $v$, then we find an expression $H_c$ through $M$ and parametre $\alpha$, and then, by using (1), an expression for $r_c$:

$$H_c^{-1} = G(\alpha^3 M)^{1/4}, \quad r_c = G(\alpha M)^{1/2}. \tag{9}$$

According (9), to a galaxy of larger mass corresponds lower value of $H_c$, which means that as massive galaxy, as at later epoch its GBR was formed, that is quite satisfactory.

For elliptic galaxies, which not rotate, there is the Faber-Jackson relationship $M = \alpha v_d^4$, similar to BTF, but for the velocity dispersion $v_d^2$, and it also coincides by the stretched potential.

Let us apply the results for single galaxy's field to groups and clusters of galaxies. Flatness of rotation curves, due to the principle of equivalence, is applicable also to a system of two galaxies of total mass $M_1 + M_2$. This binary system can be reduced to motion around the centre mass of a body of reduced mass $\mu = M_1 M_2 / (M_1 + M_2)$. Therefore, stretching of the potential can be tested also by studying of dynamics of binary systems at which rotation velocity, depending on total mass, should be practically independent on radius of the system.

If the size of a binary system of galaxies is close to the size of each of them, then part of matter falls to the centre of mass of the system by forming a bulge. In addition, the centre of mass of each of former galaxies, with stretched potential, will rotate with former velocity, being prototype of density waves transforming later to two or more spiral arms. The similar scenario can be realized in a system of three galaxies, where one of galaxies may has direct or opposite velocity, and which later transforms to a spiral galaxy with three or more spiral arms. Thus, at the stretched Newtonian potential the close systems of elliptic or irregular galaxies can transform by time to spiral galaxies. Notice, that flatness of rotation curves and large angular momenta of spiral galaxies follow from the large angular momentum of the initial binary system.

Finally, at considering of groups and clusters of galaxies in the present treatment, we find that the velocity dispersion in them will be defined also by the stretched potentials and Eqs. (7)-(8). Since the clusters are formed sufficiently later than an epoch when the stretched potential of galaxies was formed, we can take in (8) approximately $H_c \sim H_0$, which for baryonic mass of clusters gives the relationship

$$\frac{M_{cl}}{M_{gal}} \simeq \frac{v^3_{c(cl)}}{v^3_{c(gal)}}. \qquad (10)$$

At $v_{c(gal)} \sim 200 \, km \times s^{-1}$ for $M_{gal} \sim 10^{11} M_\odot$, in a cluster with velocity dispersion $v_{c(cl)} \sim 1000 \, km \times s^{-1}$ Eq. (10) gives $M_{cl} \sim 10^{14} M_\odot$, which coincides by observational estimations.

Thus, the stretched NP, falling by distance more slowly, explains practically all unusual effects of galactic astrophysics ascribed to dark matter and may be taken as a starting point for investigations of structure formation.

## References

1. J. Beringer et al. (PDG), Phys. Rev. **D86**, 010001 (2012).
2. M. Carrera and D. Giulini, Rev. Mod. Phys. **82**, 169 (2010).